\documentclass[12pt,preprint]{aastex}
\usepackage{epsfig}
\usepackage{lscape}

\received{}
\accepted{}
\shorttitle{IRAS~16547-4247}
\shortauthors{Franco-Hern\'andez et al.}

\newcommand{\etal}{\textsl{et~al.}}
\newcommand{\ala}[1]{\ensuremath{^{#1}}}
\newcommand{\rmaa}{\textsl{Rev. Mex. Astron. Astrophys. Conf. Ser.}}

\begin{document}

\title{The Rotating Molecular Structures and the Ionized Outflow Associated with IRAS 16547-4247}

\author{Ramiro Franco-Hern\'andez\altaffilmark{1,2}, James M. Moran\altaffilmark{1}, Luis
F. Rodr\'\i guez\altaffilmark{2}, and Guido
Garay\altaffilmark{3}}  
\altaffiltext{1}{Harvard-Smithsonian Center for Astrophysics,
    60 Garden Street, Cambridge, MA 02138. rfranco@cfa.harvard.edu}
\altaffiltext{2}{Centro de Radioastronom\'\i a y Astrof\'\i sica, UNAM,
Apartado Postal 3-72 (Xangari), 58089 Morelia, Michoac\'an, M\'exico.}
\altaffiltext{3}{Departamento de Astronom\'\i a, Universidad de Chile,
Camino el Observatorio 1515, Las Condes, Santiago, Chile.}

\begin{abstract}
We present VLA 1.3 cm  radio continuum and  water maser observations as
well as SMA  SO$_2$ (226.300 GHz) and 1.3 mm dust continuum
observations toward the massive star formation region
IRAS~16547-4247. We find evidence of multiple sources in the 
central part of the   
region.  There is evidence of a rotating structure associated with the   
most massive of these sources, traced  at small scales ($\sim$50 AU) 
by the water masers. At large scales ($\sim$1000 AU) we find a velocity 
gradient in  
the SO$_2$ molecular emission with a barely resolved structure that can be
modeled as a rotating ring or two separate objects. 
The velocity gradients of the masers and of the molecular emission have the same 
sense and may trace the same structure at different size scales. 
The position angles of the structures associated with the velocity gradients are 
roughly perpendicular to the outflow axis observed in radio
continuum and several molecular tracers. We estimate the mass of the
most massive central source to be around 30 solar masses from the
velocity gradient in the water 
maser emission. The main source of error in this
estimate is the radius of the rotating
structure. We also  
find  water masers that are  associated with the large scale
molecular outflow of the system, as well as water masers that are 
associated with  other sources in the region. 
Our results suggest that the formation of this source, one of the most luminous protostars
or protostellar clusters
known, is taking place with the presence of ionized jets and disk-like structures. 

\end{abstract}


\section{Introduction}

Our present understanding of star formation is primarily based on  
observations of the relatively abundant low-mass stars. The
theoretical framework for star formation (Shu et al. 1987; 1993, see also
McKee \& Ostriker 2007) has been 
successful in explaining  the processes that occur in the formation of
these low mass stars, processes that are inferred from multiwavelength observations
(e.g., Lada 1991, Evans 1999). Key ingredients in this scenario are the  
presence of a central protostar accreting from a circumstellar disk
that is surrounded by an 
infalling envelope of dust and gas, as well as the presence of ionized
jets and  molecular outflows that remove angular
momentum and mechanical  energy from the accretion disk.

The applicability of this paradigm to the formation of massive stars remains 
unproven. It is possible that massive stars are formed by processes that are 
radically different from those that produce low-mass stars, such as by the 
merging of lower mass protostars (Bonnell, Bate, \& Zinnecker 1998).
The role of the coalescence (Stahler, Palla, \& Ho 2000) and accretion
(Osorio, Lizano, \& D'Alessio 1999, McKee \& Tan 2002) processes in the 
assembling of a massive star is still under debate. If massive O stars are 
formed by accretion we expect that disks and jets will be present 
during their earliest stages of evolution. On the other hand, if they are
formed through coalescence  of lower-mass stars then neither disks nor
jets are expected since they would be disrupted during the merging
process. For a recent review on the competing ideas to explain massive
star formation see Zinnecker \& Yorke (2007).

Water masers have been observed in association with massive  star
formation regions and even though they have been studied for four decades
(Cheung et al. 1969), their nature is still not fully
understood. There has been a lot of discussion  
on where in the star formation region the physical conditions match  those
necessary for the excitation of the water  masers.    
One idea is that the masers originate in a layer between the
ionization and shock fronts in the expanding HII 
regions (e.g. Elitzur 1992). Another is that  they are formed at the
interface of the molecular material with the jets
and outflows emanating  from the forming stars
(e.g. Fuyura \etal\ 1999). 
Some observations also suggest that the water masers trace
circumstellar disks (e.g. Torrelles \etal\ 2002).

IRAS~16547-4247 is a luminous infrared source (bolometric luminosity of 
$6.2\times10^4$ $L_\odot$), located at a distance of 2.9 $\pm$ 0.6
kpc (Rodriguez \etal\ 2008). It is thought to be 
associated with a massive young stellar object. Low angular resolution 
($\sim 24{''}$) millimeter wavelength observations show that the IRAS source is
associated  with an isolated and dense molecular core with a mass of
$1.3\times10^3$ $M_\odot$  and a radius of 0.2 pc (Garay et al. 2003).
Very Large Array (VLA) and Australia Telescope Compact Array (ATCA) 
interferometric observations of radio continuum at centimeter wavelengths
show the presence of a thermal  radio jet, located at the center of
the core, and two lobes aligned and  symmetrically separated from the
jet by $\sim 10{''}$  or $\sim$0.14 pc at a P.A. of $~163^{\circ}$
(Garay et al. 2003; Rodr\'\i guez et al. 2005). In addition,
observations of H$_2$  emission at 
2.12 $\mu$m reveal a chain of knots of shock-excited
gas extending over 1.5 pc, and 11.9 $\mu$m continuum observations show
a compact  object ($\leq 0\rlap.{''}8$)
at the position of the jet (Brooks et al. 2003). The large scale H$_2$ 
flow and the radio jet are closely aligned, suggesting that these phenomena 
trace the outflowing gas at different distances from the forming star.

More recently Garay \etal\ (2007) observed the  CO
$J=3\rightarrow2$ transition  toward this source. Their
observations show the presence of a  collimated bipolar outflow
roughly oriented in the north-south direction.  The
position angle of the molecular outflow (P.A. = $~174^{\circ}$) is
slightly different from the position angle 
of the radio jet (P.A. = $~167^{\circ}$), suggesting that
the jet axis is precessing.
Additional evidence for precession has been presented
by Rodr\'\i guez et al. (2008). Garay \etal\ (2007) estimated  an inclination
for the outflow of $i=84\pm2^{\circ}$ from comparison of the
position-velocity diagram of the outflowing CO with the biconical models of
Cabrit et al. (1988). The momentum and energy
parameters they  derived  are consistent with a  massive young star
driving the outflow.  Also Garay \etal\
observed a strong signature of large scale  infall
motions and derived an infall speed of $\sim$1 km s$^{-1}$
and a mass infall rate  of $\sim1\times10^{-2} M_{\odot}$ yr$^{-1}$ at a radius
of $\sim$0.2 pc. 

Water masers were detected toward IRAS~16547-4247 by
Batchelor et~al. (1980) with the Parkes radio telescope. They  reported a
flux of 100 Jy and a v$_{LSR}=-34$ km s\ala{-1} for the maser
emission.  Later Forster and Caswell (1989) made a brief   VLA
observation  of IRAS~16547-4247 for water masers.  The angular
resolution on their  observations  was  enough to separate the masers
into a few  groups located near the central radio continuum
source. However, due to their short integration time  ($\sim$8 min)
and narrow bandwidth, the radio continuum was  not detected.

Here we present
water maser and radio  continuum observations
at 1.3 cm made  with the VLA of the NRAO\footnote{The National Radio
Astronomy Observatory is operated by Associated Universities
Inc. under cooperative agreement with the National Science Foundation.}.
We also present new observations made 
with the SMA\footnote{The Submillimeter Array is a joint project between the 
Smithsonian Astrophysical Observatory and the Academia Sinica Institute of Astronomy 
and Astrophysics and is funded by the Smithsonian Institution and the 
Academia Sinica.} of the SO$_2$ $14(3,11)-14(2,12)$ 
transition at 226.300 GHz, as well as the 1.3 mm dust continuum 
emission.  
In section \ref{observations} we describe the observations, while in
section \ref{results} we present our results. In section
\ref{discussion} we discuss the data, 
and in section \ref{conclusions} we summarize our conclusions.

\section{Observations}\label{observations}

The VLA observations were taken on 2007 June 7, using 
the correlator in the observing mode 2AB. This allowed us to observe
simultaneously the  radio continuum and spectral line emission.  The  total
bandwidth for the continuum was  25 MHz. For the  spectral line observations we
had a total of 64 
channels with a resolution of 97.6 kHz or 1.3 km s\ala{-1}. These
observations were taken in the  A configuration, resulting in a beam with
a FWHM of $0\rlap.{''}31\times 0\rlap.{''}07$; P.A.=0.8$^{\circ}$
for ROBUST = 0 weighting (Briggs 1995). The
data were calibrated using the standard high frequency calibration
procedures for the VLA as described in the AIPS Cookbook. The phase 
calibrator was  J1717-398 which had a bootstrapped flux of 0.30 $\pm$
0.01 Jy. A line channel with bright water maser emission was used
to self-calibrate the line data and cross-calibrate the continuum
data.  

The SMA observations were taken on 2005 April 30. The array was in the
extended configuration with a beam of FWHM of
$2\rlap.{''}5\times0\rlap.{''}97$; P.A.=8$^{\circ}$ (for a weighting with
ROBUST = 0), while the
spectral resolution was  1.1 km s$^{-1}$. The calibration was performed
using MIR and MIRIAD.  Callisto  was used for flux calibration and
3C279  for bandpass calibration. The phase calibrator was 1745-290, with
a bootstrapped flux of 3.15 Jy with a 20\% accuracy. The continuum data at 1.3
mm were processed in MIRIAD using the task UVLIN in the lower sideband
with  2 GHz bandwidth centered at 217.1 GHz.  The continuum was then
self-calibrated in phase  using CALIB in AIPS. The resulting
calibration was then applied to the SO$_2$ line data.
 
\section{Results}\label{results}

\subsection{1.3 cm  VLA observations}

Since IRAS~16547-4247 is a rather southern source (-42$^\circ$ in
declination) the  VLA had  to observe  through a large  air mass
during the whole observation run. This adversely
affected the phase stability especially in poor weather. In the VLA observations
presented by Rodr\'\i guez \etal\ (2005) the weather
was not good and they used a self calibration
technique to obtain high dynamic range  maps. Although the
self calibration largely  improved the dynamic range, it 
did not improve
the accuracy of the position of the source.  

We made a special effort to determine the source position accurately. To
obtain a better estimate for the  position of the 
radio continuum we looked at the positions obtained from observations
made in good weather where we could measure the position of the central bright radio
continuum source. Rodr\'\i guez \etal\ (2005)
reported positions derived from ATCA obserations in 2003 February
of $\alpha(2000) = 16^h~ 58^m ~ 17\rlap.^s216, \delta(2000) =  -42^\circ~ 52'~
07\rlap.{''}64$  at 6 cm and  $\alpha(2000) = 16^h~ 58^m~ 
17\rlap.^s210, \delta(2000) = -42^\circ ~ 52' ~ 07\rlap.{''}48$ at 3.6
cm. The phase calibrator used in these
observations was 1616-52. From the VLA 
archives there is data taken on 1993 January. The position
measured from this epoch is $\alpha(2000) = 16^h~ 58^m ~ 17\rlap.^s246, \delta(2000) =
-42^\circ ~ 52' ~ 07\rlap.{''}88$ at 3.6 cm with the phase calibrator
1626-298. We have a position from the new VLA observations from 2006 May 31 and June 8
(Rodr\'\i guez et al. 2008),
that, after being concatenated, give the position $\alpha(2000) = 16^h~ 58^m ~ 
17\rlap.^s2093, 
\delta(2000) =  -42^\circ ~ 52' ~ 07\rlap.{''}150$ at 3.6 cm with the
phase calibrator 1626-298. Since the
newest VLA  data from 2006 has much more integration time than the
snapshot from the archive we will just use the position from
the former data, and from the two
ATCA positions we take the position at 3.6 cm because the 6 cm data
has lower angular resolution. 
We adopt as the final position estimate for the central bright radio
continuum source at centimeter wavelengths the average  of the ATCA
position at 3.6 cm (Rodr\'\i guez \etal\ 2005) and the most recent (2006) VLA position
(Rodr\'\i guez \etal\ 2008). This gives $\alpha(2000) = 16^h~ 58^m ~
17\rlap.^s210 \pm 0\rlap.^s001, \delta(2000) = -42^{\circ} ~ 52' ~ 07\rlap.{''}32 
\pm 0\rlap.{''}17$,
where the errors quoted are one half the difference between the two
observations used.  

\begin{figure}[ht]
\centering
\includegraphics[width=0.6\textwidth,angle=0]{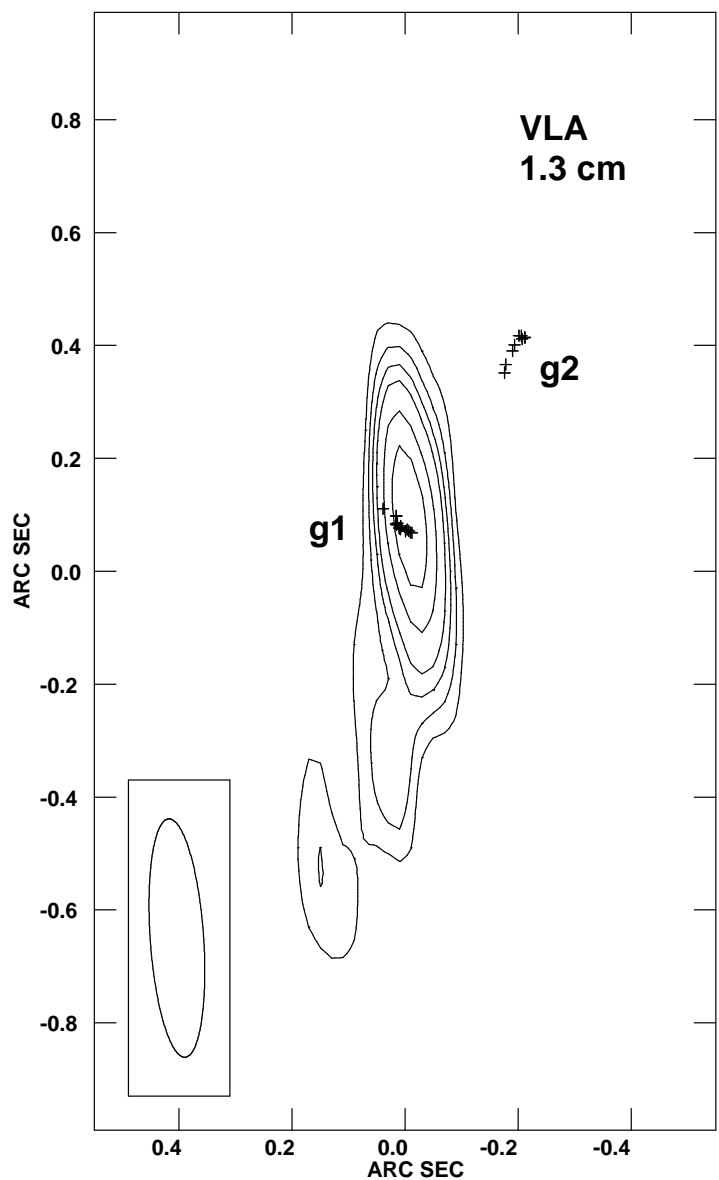}
\caption{\label{i16547kcont} VLA 1.3 cm continuum emission from the 
IRAS~16547-4247 central source. The image was made with ROBUST = 5 weighting. 
Contours are -4, -3, 3, 4, 5, 6, 8, and 10 times 0.42 mJy
beam$^{-1}$, the rms noise of the map. The half power contour of the
beam ($0\rlap.{''}43 \times 0\rlap.{''}09; 4^\circ$), is shown in the bottom left corner.
The peak position adopted for this source is
$\alpha(2000) = 16^h~ 58^m ~
17\rlap.^s210, \delta(2000) = -42^{\circ} ~ 52' ~ 07\rlap.{''}32$.
The main component is marginally resolved and elongated along PA of $177^\circ$. The small crosses
mark the positions of the water masers in the compact structures located in groups
\textbf{g1} and \textbf{g2}. } 
 
\end{figure}

We used self-calibration techniques to obtain a final
image for the  new observations at 1.3 cm presented here.
We asssume that the position of the peak 1.3 cm emission coincides with
our adopted position.
Since the water masers were observed
simultaneously with the continuum, their  positions are also
referenced to the same coordinates. It is worthwhile to note that the
relative positions between the water masers and the 1.3 cm continuum
are very accurate and independent of the
absolute position adopted in the self-calibration. 
The image of the 1.3 cm continuum emission is shown in Fig.~\ref{i16547kcont}.
The emission is dominated by the thermal jet, for which we derive,
from a Gaussian ellipsoid fit made using the task JMFIT
of AIPS, a total flux density of 10.9$\pm$1.3 mJy 
and deconvolved angular dimensions of $0\rlap.{''}55 \pm 0\rlap.{''}08 \times
0\rlap.{''}13 \pm 0\rlap.{''}03$ with PA of $177^\circ \pm 4^\circ$.
The position angle of the major axis is consistent with that derived at longer
wavelenghts (Garay et al. 2003, Rodr\'\i guez et al. 2005, 2008), indicating that at
7 mm we are tracing the ionized jet.
The image shows two marginal (4-$\sigma$) components to the south
of the jet that require confirmation in deeper images. 

\begin{figure}[ht]
\centering
\includegraphics[width=0.8\textwidth]{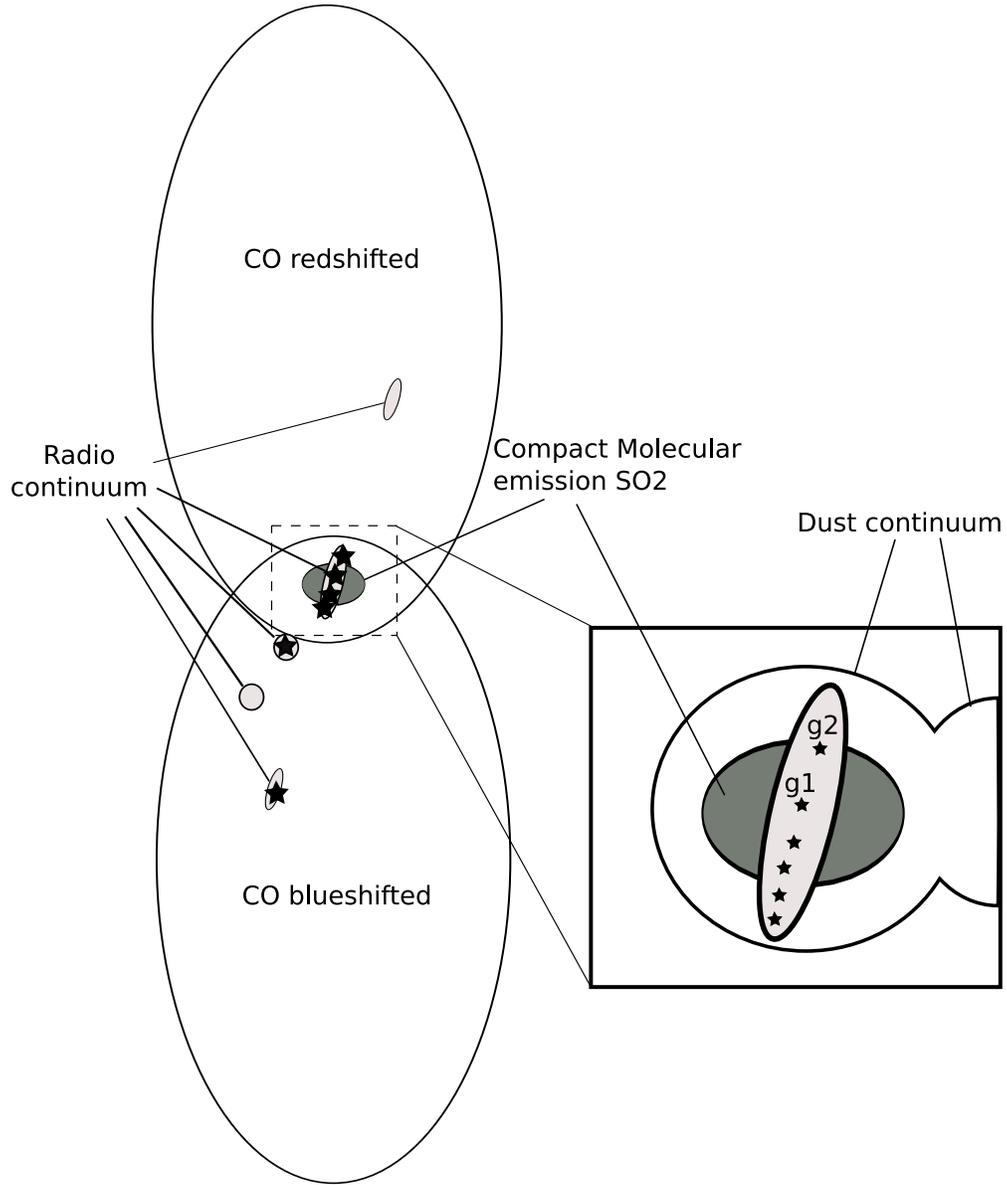}
\caption{\label{cartoon} Cartoon (not to scale) showing the different features
and components of IRAS 16547-4247. The stars mark the positions of
water maser groups.}
\end{figure}

The distribution of the main features of the
IRAS~16547-4247 region, including that of
the different water maser groups is shown in Fig.~\ref{cartoon}. The
parameters of the individual maser spots
are listed in Table~\ref{h2omasertable}. It is clear that there are
several groups of masers associated with different radio continuum
sources seen in the  3.6 cm continuum map from Rodr\'\i guez \etal\
(2005). These groups are separated with horizontal
lines in Table.~\ref{h2omasertable}. 
The strongest masers, in the group marked as \textbf{g1},
are associated with the central and
brightest radio continuum source.
Their radial velocities range from -25.3 km s\ala{-1} to -54.2 km
s\ala{-1}.
Most of the masers in the
\textbf{g1} group form a compact structure extending in the east-west
direction (see Fig.~\ref{masers1}). 
There is another clear group of masers denoted \textbf{g2} in
Fig.~\ref{cartoon}. This group is located 
to the north-west of the radio continuum peak and the maser radial
velocities extend from -30.5  km 
s\ala{-1} to -42.4 km s\ala{-1}. Most of the masers in the
\textbf{g2} group form a compact structure that is
shown in Fig.~\ref{masers2}. We will discuss these two compact structures below. 

\subsection{SMA 1.3 mm continuum observations}

The SMA dust continuum map presented in Fig.~\ref{dust1} clearly shows
that the emission at 1.3 mm exhibits a morphology that looks like two 
partially resolved
sources separated by about $2\rlap.{''}0$.
We fitted the emission to two gaussian ellipsoids with the AIPS
task JMFIT. The parameters resulting from the fit 
(positions, flux densities, deconvolved angular sizes and position
angles) are given in Table \ref{1mmcont}.
From the fit we find that, as expected from Figure 5,
the eastern component is
the brightest in the continuum. This
component is also detected in several molecular lines
(Franco-Hern\'andez \etal\, in preparation). The total dust continuum
flux is 1.03 Jy, split in  0.81 and 0.22 Jy for the eastern and
western components, respectively.  Following Chini \etal\ (1987),
assuming a constant temperature of 300  K  (appropriate for dust at
1000 AU from a $6\times10^4$ $L_{\odot}$ source), optically-thin
emission with a dust to gas ratio of 0.01, and a dust mass opacity of
1 cm$^2$ g$^{-1}$ at 1.3 mm, we calculate a 
total gas mass in the double dust continuum source of $\sim$6 $M_\odot$, split into 4.7 and
1.3 $M_\odot$ for the strong and weak components, respectively.   

The western component is unresolved, while the
deconvolved angular sizes (FWHM) obtained from the fit for the eastern component
are $1\rlap.{''}34\times 0\rlap.{''}84$; P.A.= 107$^{\circ}$.
The stronger 1.3 mm component can be interpreted as a flattened
structure, approximately in the east-west direction with a size of $\sim 3500$ AU.
This structure may be tracing a large disk around the central
source or sources.  

It is interesting to note that the west (secondary) 1.3 mm component appears
to be a source different from the centimeter source D (Rodr\'\i guez et al.
2008), that also appears
to the west of the main source. While the west 1.3 mm source is displaced by
$1\rlap.{''}94 \pm 0\rlap.{''}15$ from the main source, the centimeter
source D is displaced by
$1\rlap.{''}02 \pm 0\rlap.{''}03$ from the main source.
We conclude that there are at least three sources within 2${''}$ of the
main source: the main source itself, the centimeter source D, and
the west 1.3 mm source reported here.

Finally, we note that the centimeter position
adopted for the main source,
$\alpha(2000) = 16^h~ 58^m ~ 
17\rlap.^{s}210 \pm 0\rlap.^{s}001; \delta(2000) = -42^{\circ}~ 52'~ 07\rlap.{''}32 
\pm 0\rlap.{''}17$,
and discussed in Section 3.1, does not coincide within the errors
with the position
obtained from the 1.3 mm image (see Table 2),
$\alpha(2000) = 16^h~ 58^m ~
17\rlap.^{s}247 \pm 0\rlap.^{s}002; \delta(2000) = -42^{\circ}~ 52'~ 08\rlap.{''}09
\pm 0\rlap.{''}04$ and that these positions differ by
$\sim 0\rlap.{''}87$.
We tentatively attribute this discrepancy to the use
of different phase calibrators. 
To facilitate modelling of the region we assume
that the SMA 1.3 mm peak coincides
with the VLA/ATCA 3.6 cm peak. The same offset is applied to the SMA
molecular data.

\subsection{SMA SO$_2$  observations}

We detected strong SO$_2$ emission
toward the dominant eastern dust component, 
but not toward the weaker western component. 
Fig.~\ref{dust}  shows the first moment of the SO$_2$ 226.300
GHz transition together with the 1.3 mm continuum emission,
both overlaid on the VLA 3.6 cm emission. The SO$_2$ emission 
comes from a structure of $\sim$1500 AU in
radius.  We measured the peak position in each velocity  channel of the
SO$_2$ emission. In Table \ref{so2table} we list these  positions and
they are plotted in Fig.~\ref{so2}. In this plot it can be seen that the
velocity gradient is roughly perpendicular to  the direction of the outflow
observed in CO (Garay \etal\ 2007) and the radio jet (Garay \etal\
2003, Rodr\'\i guez \etal\ 2005, this paper). 

\section{Discussion}\label{discussion}

\subsection{Water masers}

\begin{figure}[ht]
\centering
\includegraphics[width=0.8\textwidth]{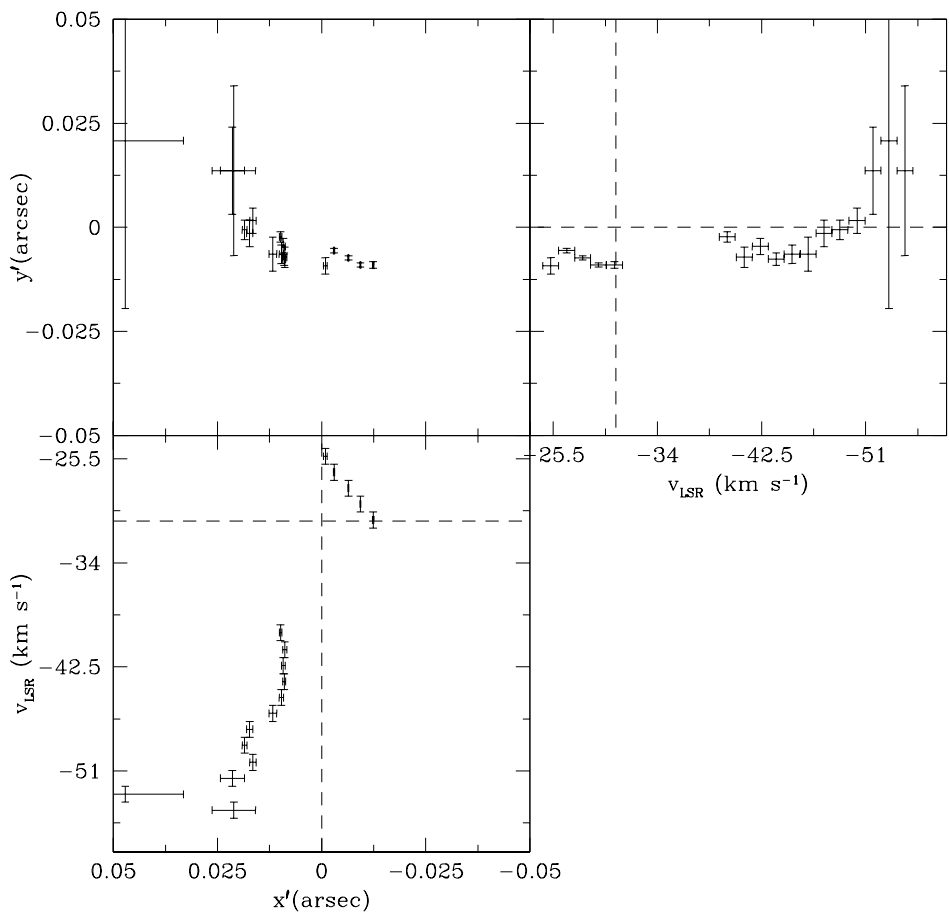}
\caption{\label{masers1} Distribution in position and radial velocity of
the compact structure of masers in 
\textbf{g1} (Fig.~\ref{cartoon}). The upper left panel shows the
positions of the masers with respect to the peak of the radio
continuum at 1.3 cm. The upper right and lower panels are the 
position-velocity diagrams for the same masers. The broken lines show the
position of the radio continuum and the LSR velocity of the source in
their respective axes. The positions have been rotated $13^\circ$ 
counterclockwise to make the ordinate parallel to the ionized jet. The errors
in position shown are ten times larger than the real values.}  
\end{figure}

\begin{figure}[ht]
\centering
\includegraphics[width=0.8\textwidth]{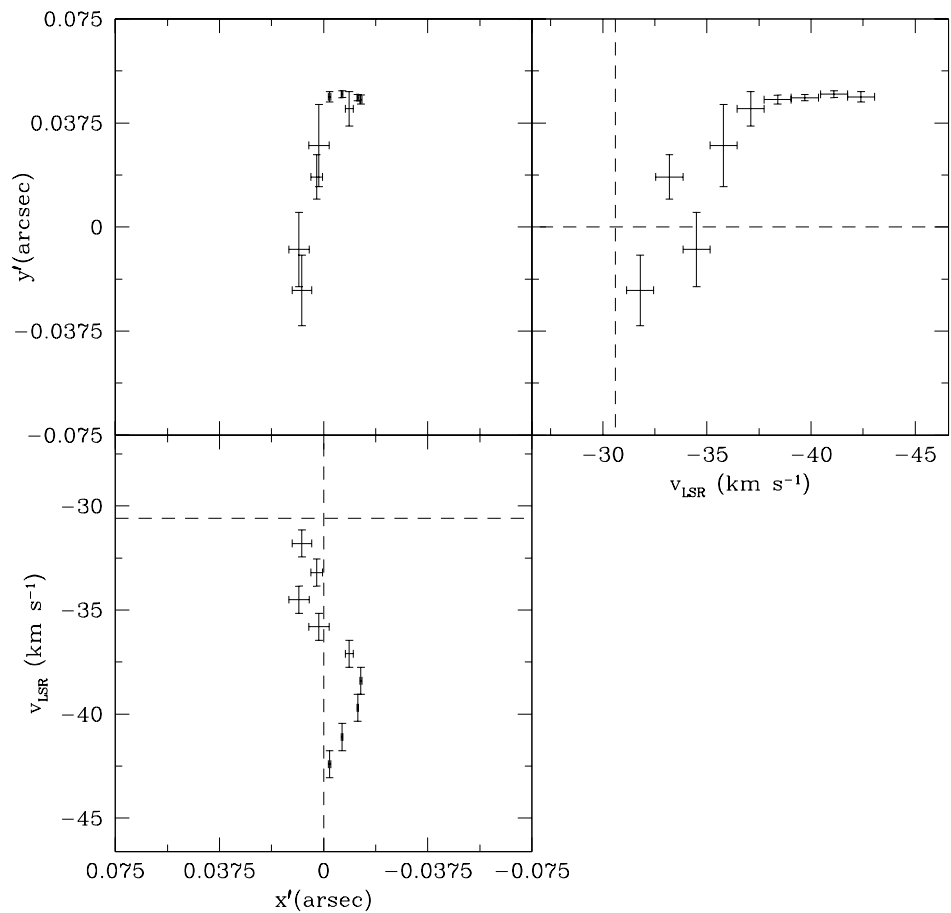}
\caption{\label{masers2} Distribution in position and radial velocity of
the compact structure of masers in 
\textbf{g2} (Fig.~\ref{cartoon}). Panels are as in Fig.~\ref{masers1}. A
clear gradient can be seen to the north of the peak of
radio continuum. Reference position is $\alpha(J2000) = 16^h~ 58^m ~ 07\rlap.^s1930,
\delta(J2000) =
-42^{\circ} ~ 52' ~ 7\rlap.{''}028$. The positions have been rotated $13^\circ$
counterclockwise to make the ordinate parallel to the ionized jet. The errors
in position shown are ten times larger than the real values.} 
\end{figure}

\begin{figure}[ht]
\centering
\includegraphics[width=0.8\textwidth,angle=0]{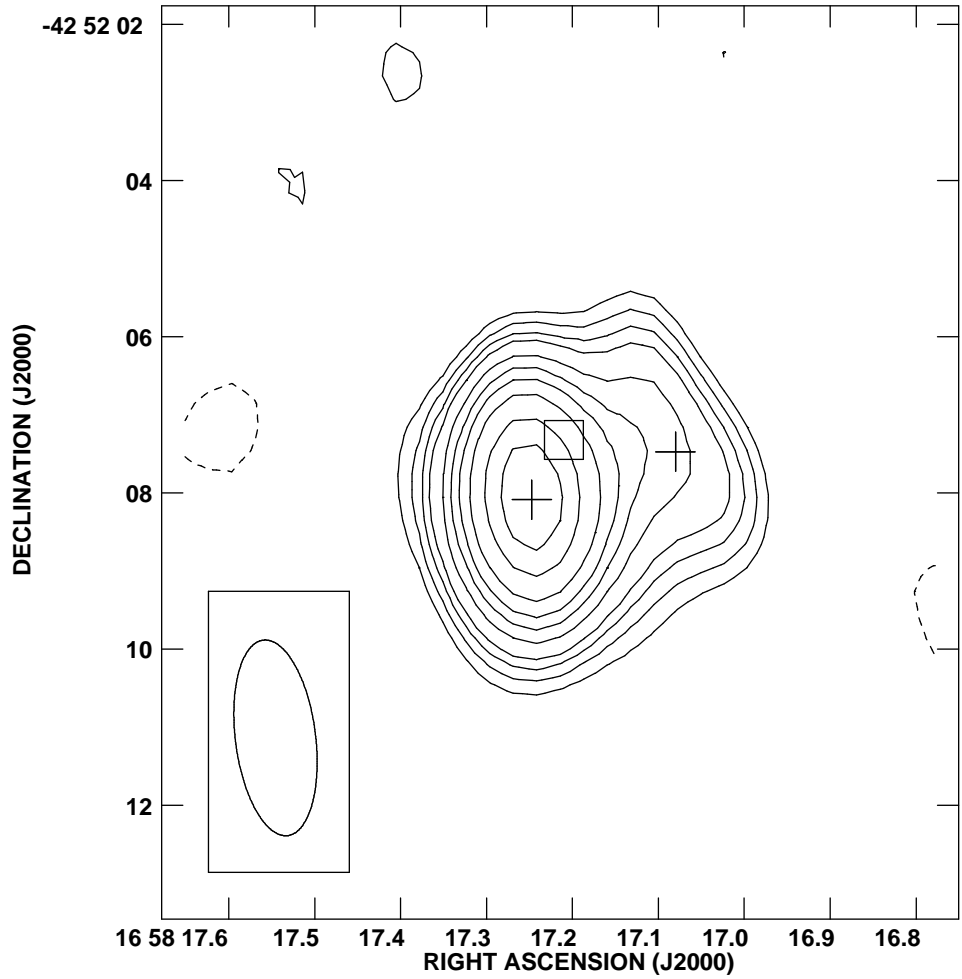}
\caption{\label{dust1} SMA 1.3 mm dust continuum emission from the 
IRAS~16547-4247 central source. 
Contours are -4, -3, 3, 4, 5, 6, 8, 10, 12, 15, 20 and 25 times 16 mJy
beam$^{-1}$, the rms noise of the map. The crosses mark the position of the
two components obtained from the fit to the image, whose parameters
are given in Table 2. The square box marks the position finally adopted
for the peak 1.3 mm emission (see discussion in text).}
 
\end{figure}

\begin{figure}
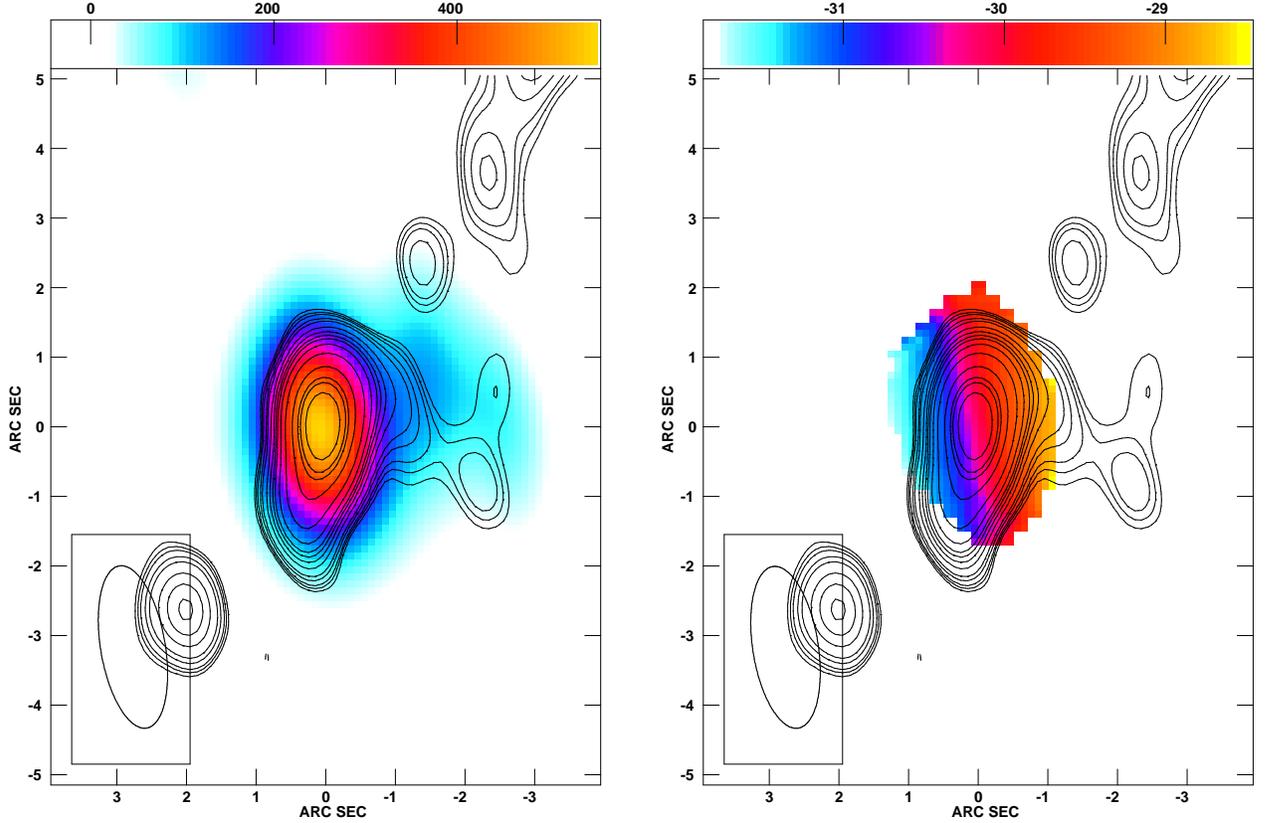

\centering
\begin{tabular}{cc}
\includegraphics[width=0.5\textwidth,angle=0]{f6a.eps}&
\includegraphics[width=0.5\textwidth,angle=0]{f6b.eps}
\end{tabular}
\caption{\label{dust} The left panel shows the 1.3 mm
dust continuum emission  in color and the 3.6 cm free-free radio
continuum in contours. The color bar at the 
top shows the color coding for the 230 GHz flux density in
mJy beam$^{-1}$. Contour levels for the 3.6 cm
emission are -5, 5, 8, 10, 15, 20, 40, 60, 80, 100, 140,
180 times the rms noise level of 30 $\mu$Jy beam$^{-1}$. The 3.6 cm
source at the lower left corner of the image 
most probably traces an independent star
and is not associated with
the jet (Rodr\'\i guez et al. 2005). The right panel shows   
the first moment map  of the 226.300 GHz SO$_2$ transition in color
and the radio continuum emission at 3.6 cm  with the same contours as
in the left panel (Rodr\'\i guez et al. 2005). The color bar at the
top shows the color coding for the LSR velocity of the gas
in km s$^{-1}$. Note that the  western component is not detected in
the SO$_2$ emission (right panel). The synthesized beam for the 3.6 cm
data is shown in the bottom left corner of each panel. The synthesized
beam for the 230 GHz data is 2.3 $\times$ 0.8; PA = 11$^{\circ}$ }
\end{figure}

In Fig. \ref{masers1} we show the relative  positions of the water
masers in the compact structure in group \textbf{g1} with respect to the radio continuum at 1.3
cm. It can be seen  that the  position of the peak of the radio
continuum lies close to the position of the structure.
From  the lower left panel of Fig. \ref{masers1} it is tempting
to infer that the LSR 
velocity for the source is around -42 km s\ala{-1} since this is the
central velocity for this group of  masers and would make them look
like a complete  rotating ring with masers red and blueshifted
with respect to this velocity. However, this velocity is 12 km s$^{-1}$
blueshifted with  respect to  the LSR velocity measured from the
molecular lines. This difference is large compared with
the radial dispersion velocity in  young 
stellar clusters which is expected to be only a few km s$^{-1}$
(e.g. $\sim 2$ km s$^{-1}$ for the center of the Orion Nebula, Jones
\& Walker, 1988, as well as for Cyg OB2, Kiminki \etal\ 2007). 

If  the velocity  center of the \textbf{g1} compact structure  is
at the LSR velocity of -30.6 km s$^{-1}$ we can still interpret it
as a rotating torus around the central 
source.  However, in this case we  have emission coming from masers mainly
blueshifted with respect to the LSR velocity. The lack or weak
emission from  one side of the disk (blueshifted or redshifted) has
been observed in other sources (Val'Tts \etal\ 2005) and it  is
probably related to different excitation conditions in different parts
of the ring.  Assuming the central source is at the position marked by the
peak of the radio continuum and that the maser emission is coming from
a ring in a Keplerian orbit, we can estimate
the central mass. 

We first obtain the gradient in velocity adjusting a straight line to
the masers in the velocity range  from -38.4
to -54.2 km s\ala{-1}. For this fit we discard the maser at  -52.9 km
s\ala{-1} which is 
very noisy and seems to depart from the rest of the features.
We obtain a slope of $d V/d \theta$ = -812 $\pm$
71 km s$^{-1}$ arcsec$^{-1}$ and an  intercept of -35 $\pm$ 1 km
s$^{-1}$. In the previous expression $V$ is
the observed radial velocity of the maser at a measured projected angular distance
$\theta$ from the source (this distance is the component measured in the
coordinate defined by the semimajor axis of the ring). We also need to know the outer radius
of the ring. However, all we can obtain is a lower limit for the outer radius
since  we cannot tell  if the last maser we see is coming from the
edge of the ring or if there is material external to this
apparent last maser.  Another uncertainty comes from the inclination of
the ring with respect to the line of sight. We will use the
inclination given by Garay \etal\ (2007), i.e. the ring is almost
perpendicular to the plane 
of the sky. With this in mind, we 
define the radius of the disk as the angular distance from the central source
given by the fit to the slope for the highest
radial velocity detected. For a velocity of -54.2 km s\ala{-1},
we obtain $\theta$ = 0.024$\pm$0.002 arcsec.
The expression for the mass is
\begin{equation}\label{eq1}
M_\star=\frac{\theta^3 D}{G \sin^{2} i}\Big(\frac{\mbox{d}V}{\mbox{d}\theta}\Big)^2,
\end{equation}  
where $\theta$ is the ring radius, $D$ is the distance to
the source, and $i$ is the inclination angle.
Introducing the
measured values we get a mass 
\begin{equation}\label{eq2}
M_\star=30 \pm 9~ M_{\odot}
\left(\frac{D}{2.9~\mbox{kpc}}\right).
\end{equation}
The error comes mainly from
the uncertainties in the radius of the ring  and  the
velocity gradient.

We now briefly discuss the \textbf{g2} compact structure of
water masers. In Fig. \ref{masers2} we 
show the relative  positions of the water
masers in this structure. They distribute in  an elongated shape, approximately following the
jet direction in that region. This suggests that these masers are tracing the outflow.
However, the masers are blueshifted with respect to the systemic velocity of
the region, while the CO outflow is redshifted. There is also a strong velocity gradient
in the northernmost part of the structure, suggesting interaction with ambient
material. 

\subsection{SO$_2$}

Sulfur-bearing molecules have been studied through   millimeter and submillimeter
observations of star formation regions. They could account for a significant
fraction of the total
flux coming from molecules in these regions (Schilke \etal\ 1997). 
The chemistry of these molecules has been modeled and its  relative abundances 
have been  explored as a possible clock to measure the 
evolutionary state in low and high mass  star formation regions 
(Buckle \& Fuller 2003, Hatchell \etal\ 2003, van der Tak \etal\ 2003). 
In the chemical models the SO$_2$  is the product of reactions  from molecules 
that  have been evaporated off the dust grains surfaces (Charnley 1997). 
There is evidence  that  the emission originates  in outflows  
(e.g. Codella \etal\ 2005, Schilke \etal\ 1997) but also in what look like 
rotating structures around some 
sources (Cep A HW2; Jim\'enez-Serra \etal\ 2007, AFGL 2591; van der Tak \etal\ 2006). 
SO$_2$ also has been observed to trace infall motions (e.g. W51; Sollins \etal\ 2004). 

\begin{figure}[ht]
\centering
\includegraphics[width=0.8\textwidth]{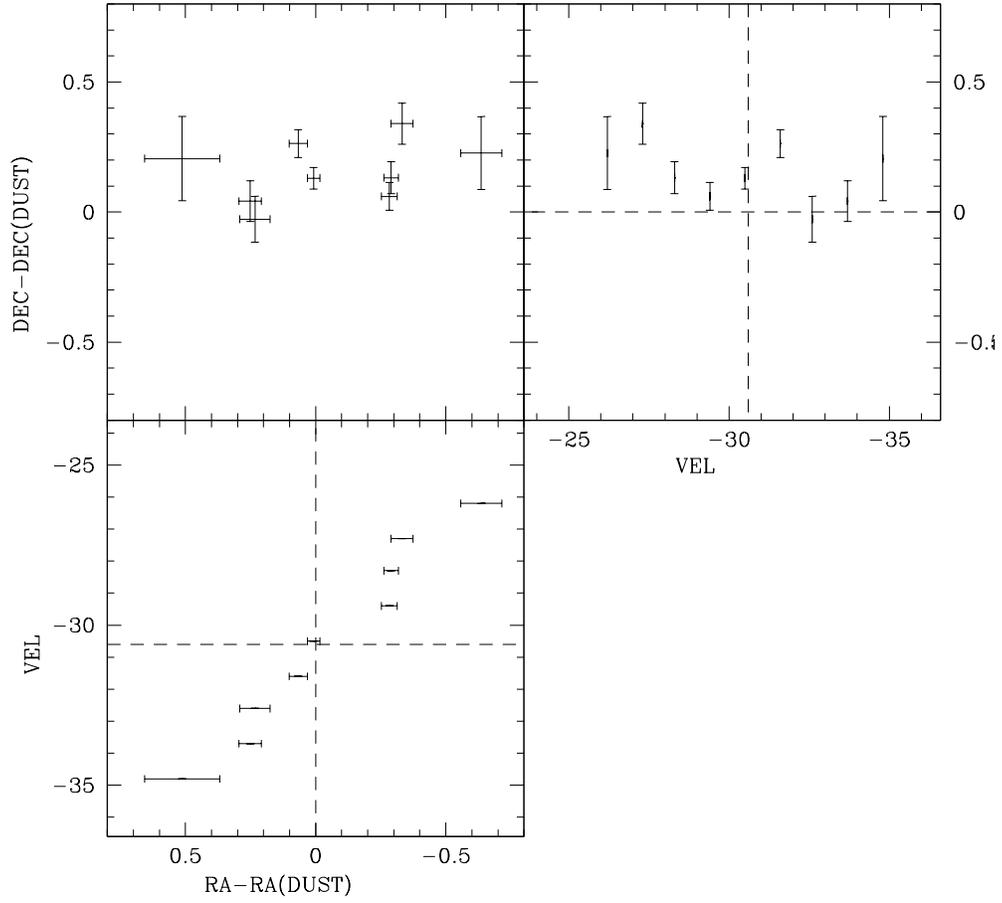}
\caption{\label{so2}The upper left panel shows the positions of the SO$_2$
(226.300 GHz) emission peaks, measured from each velocity
channel and listed in Table \ref{so2table}. The positions are offsets
with respect to the peak of the dust continuum.
The upper right and lower 
panels are the position velocity diagrams for the DEC and RA directions,
respectively.}  
\end{figure}

From the velocity gradient observed in the  SO$_2$ transitions we
obtain an independent estimate of the mass. In Fig. \ref{so2} we
can see the position velocity diagram showing the position of the
emission peak in each velocity channel. This can be
interpreted also as gas in a Keplerian motion in a ring around the
central source. The 
estimate of the size of such a ring will have the same uncertainties as
described above for the water masers and we will have again a lower
limit for the mass estimate. For these data we get a velocity gradient
of -8.8 $\pm$ 0.8 km s$^{-1}$ arcsec$^{-1}$, and an intercept of
-31.0 $\pm$ 0.3 km s$^{-1}$. 
The velocity of the most blueshifted detectable SO$_2$ emission is
-34.8 km s$^{-1}$. For this velocity,
we obtain $\theta$ = 0.43$\pm$0.05 arcsec. 
Using eq.~(\ref{eq1})  we obtain an estimate of the
mass given by: 
\begin{equation}\label{eq3}
M_\star=22 \pm 8~ M_{\odot}
\left(\frac{D}{2.9~\mbox{kpc}}\right),
\end{equation}
where the errors are estimated in the same way that as for the water masers. 

Then, the Keplerian masses estimated from the water masers and from the 
SO$_2$, at physical scales that differ by a factor of $\sim$20, give
consistent large masses for the central star(s). However, the SO$_2$ data set
has very limited angular resolution and we find that the apparent gradient 
seen in Fig.~\ref{so2} can also be modeled as a Gaussian two-component model.
From this fit, we find two components separated by $0\rlap.{''}74 \pm 0\rlap.{''}06$
at a position angle of $109^\circ \pm 6^\circ$, and with a velocity
difference of $2.3 \pm 0.2$ km s$^{-1}$. These two components, if assumed to
be of negligible mass and located symmetrically with respect to a central object with
all the mass (that is not directly detected in our observations), imply a Keplerian 
mass of $1.6 \pm 0.4~M_\odot$, much smaller that the mass derived
from the continuous ring model. Alternatively, we can assume that the
total mass of the system is distributed between the two components 
and in this case the Keplerian mass is $12.8 \pm 3.2~M_\odot$.
Given the limitations of the data, it is very difficult
to decide if the SO$_2$ emission is coming from a continuous, ring-like structure
or from two discrete molecular clumps.

\subsection{An accretion origin for the high luminosity?}

We have assumed that the high bolometric luminosity, $L_{bol} = 6.2\times10^4$ $L_\odot$, associated
with IRAS~16547-4247 comes mostly from an embedded, massive main sequence star.
This assumption is justified since high mass objects are expected to quickly
evolve to the main sequence, even while accreting and while they are deeply
embedded within the dusty core (Zinnecker \& Yorke 2007).

However, an alternative explanation is a lower mass object accreting at a
very high rate. In this case the system would derive most of its
luminosity from accretion. The accretion luminosity, $L_{acc}$, is

\begin{equation}\label{eq4}
L_{acc} = {{G M} \over {R}} \epsilon \dot M_i,  
\end{equation}

\noindent where $G$ is the gravitational constant, $M$ is the mass of the star,
$R$ is the radius of the star, $\dot M_i$ is the
infall rate determined from observations al large scales, and
$\epsilon$ is the fraction of this large scale infalling gas that ends
being accreted by the star. Assuming that the bolometric luminosity comes
mostly from accretion, $L_{acc} \simeq L_{bol}$, and that from the observations of Garay \etal\ (2007)
we have that
$\dot M_i = 1\times10^{-2} M_{\odot}$ yr$^{-1}$, we derive from Eqn. (4) a mass-radius
relationship that in solar units is

\begin{equation}\label{eq5}
\Biggl[{{R} \over {R_\odot}} \Biggr] = 5.2~\epsilon \Biggl[{{M} \over {M_\odot}} \Biggr].
\end{equation}
 
If $\epsilon$ is
larger than $\sim 0.2$, we obtain 
$(R/R_\odot) > (M/M_\odot)$ and since in the main sequence
we expect $(R/R_\odot) \simeq (M/M_\odot)$, one would have to conclude that
the star is not in the main sequence and most probably is a
lower mass object (than the mass value derived from assuming a main
sequence star) accreting at a
very high rate and deriving a significant fraction
of its total luminosity from accretion.
It should be noted, however, that Hosokawa 
\& Omukai (2009) have argued that massive protostars undergoing
strong accretion will have stellar radii an order of magnitude
larger than those occuring in the main sequence. Clearly, a better knowledge of $\epsilon$
and of the structure of protostars
will help restrict these possibilities and better quantify the contribution
of accretion to the total luminosity of forming massive stars.

\section{Conclusions}\label{conclusions}

Our main conclusions follow:

1) We present VLA 1.3 cm  radio continuum and  water maser observations as
well as SMA  SO$_2$ (226.300 GHz) and 1.3 mm dust continuum
observations toward the massive star formation region
IRAS~16547-4247. The 1.3 cm continuum traces the inner parts
of the thermal jet in the region. The 1.3 mm dust continuum traces
a double structure, with component separation of $\sim 2{''}$, 
with each structure probably marking the
position of a star or a stellar group.

2) Water maser emission is present in several distinct parts in the region.
The compact structure in group \textbf{g1} is closely associated with the
thermal jet and shows an alignment perpendicular
to it. The masers in this group show a velocity gradient that, if interpreted
as arising in a Keplerian ring, implies a mass of $\sim$30 $M_\odot$
for the central star(s).

3) The SO$_2$ emission arises only from the brightest 1.3 mm dust continuum
component. The line emission shows a velocity gradient that, if modeled
as a Keplerian ring, gives a mass of $\sim$20 $M_\odot$, consistent
with the mass derived from the H$_2$O masers. However, the data can also
be fitted with a two-component model that gives smaller
Keplerian masses. 

\acknowledgments

We thank an anonymous referee for comments that improved two sections of the
paper and for the suggestion of discussing the possibility of
significant accretion luminosity.
R.F.H. is grateful for support from an SAO predoctoral
fellowship.
L.F.R. acknowledges the support
of CONACyT, M\'exico and DGAPA, UNAM.
G.G. acknowledges support from CONICYT projects 
FONDAP No. 15010003 and BASAL PFB-06.

\clearpage

\clearpage

\clearpage

\clearpage

\clearpage

\clearpage

\clearpage

\clearpage

\begin{deluxetable}{rrrrrrrc}
\tablecolumns{8}
\tablewidth{0pc}
\tablecaption{Parameters of individual water maser spots \label{h2omasertable}}
\tablehead{

\colhead{V$_{\mbox{\scriptsize LSR}}$} &    \colhead{$\alpha$(J2000)}   & \colhead{$\Delta \alpha$}&    
\colhead{$\delta$(J2000)} & \colhead{$\Delta \delta$}  &\colhead{Flux} & 
\colhead{$\Delta$Flux}&\colhead{Continuum} \\ 
\colhead{(km s$^{-1}$)} &   \colhead{$16^h~ 58^m$}    & \colhead{(sec)} &  \colhead{$-42^{\circ}~ 52'$}   
& \colhead{(arcsec)} & \colhead{(Jy)} &  \colhead{(Jy)} & \colhead{Association$^{a}$} 
}
\startdata
-12.1 &  17.46426 & 0.00043   & 16.4344  &0.0018 &   0.37 & 0.01 &  S-1 \\
-13.4 &  17.46383 & 0.00029   & 16.4380  &0.0013 &   0.50 & 0.01 &  \\
-14.7 &  17.46567 & 0.00011   & 16.4222  &0.0005 &   1.79 & 0.01 &  \\ 
-16.0 &  17.46614 & 0.00006   & 16.4183  &0.0002 &   8.69 & 0.03 &  \\
-17.4 &  17.46617 & 0.00005   & 16.4183  &0.0002 &   11.94& 0.04 &  \\
-18.7 &  17.46618 & 0.00005   & 16.4185  &0.0003 &   5.17 & 0.02 &  \\
-20.0 &  17.46631 & 0.00014   & 16.4204  &0.0006 &   1.05 & 0.01 &  \\
-21.3 &  17.46628 & 0.00009   & 16.4199  &0.0004 &   1.87 & 0.01 &  \\   
-22.6 &  17.46628 & 0.00009   & 16.4200  &0.0004 &   1.62 & 0.01 &  \\
-23.9 &  17.46628 & 0.00023   & 16.4202  &0.0011 &   0.60 & 0.01 &     \\\hline
-27.9 &  17.36690 & 0.00469   & 10.1078  &0.0066 &   0.04 & 0.03 &  B    \\
-30.5 &  17.39604 & 0.00215   & 10.1549  &0.0102 &   0.50 & 0.07 &      \\
-31.8 &  17.39619 & 0.00083   & 10.1587  &0.0037 &   1.50 & 0.08 &      \\
-33.2 &  17.39632 & 0.00050   & 10.1617  &0.0023 &   1.72 & 0.06 &      \\
-34.5 &  17.39644 & 0.00090   & 10.1576  &0.0041 &   0.62 & 0.03 &      \\
-37.1 &  17.41415 & 0.00095   & 09.8128  &0.0183 &   0.43 & 0.04 &     \\
-38.4 &  17.37794 & 0.00123   & 10.0880  &0.0052 &   0.20 & 0.02 &     \\
-39.7 &  17.37802 & 0.00008   & 10.0873  &0.0004 &   2.73 & 0.01 &     \\
-41.1 &  17.37802 & 0.00004   & 10.0865  &0.0002 &   5.99 & 0.02 &      \\
-42.4 &  17.37801 & 0.00004   & 10.0862  &0.0002 &   4.27 & 0.01 &     \\
-43.7 &  17.37801 & 0.00022   & 10.0857  &0.0010 &   0.71 & 0.01 &     \\\hline
-18.7 &  17.22278 & 0.02314   & 08.4282  &0.0450 &   0.08 & 0.03 &  Jet    \\
-20.0 &  17.21856 & 0.00050   & 08.4653  &0.0021 &   0.31 & 0.01 &      \\
-21.3 &  17.21879 & 0.00030   & 08.4696  &0.0013 &   0.64 & 0.01 &      \\
-27.9 &  17.22072 & 0.00064   & 08.4688  &0.0027 &   1.85 & 0.07 &      \\
-29.2 &  17.22069 & 0.00015   & 08.4801  &0.0006 &   9.79 & 0.09 &      \\
-30.5 &  17.22080 & 0.00010   & 08.4815  &0.0005 &   12.96 &0.08 &      \\
-37.1 &  17.21056 & 0.00969   & 08.2030  &0.0106 &   1.49 & 0.13 &      \\\hline
-29.2 &  17.22741 & 0.00068   & 07.9606  &0.0030 &   2.66 & 0.10 &  Jet   \\ 
-30.5 &  17.22792 & 0.00003   & 07.9472  &0.0001 &   36.83 &0.07 &      \\
-31.8 &  17.22794 & 0.00002   & 07.9463  &0.0001 &   68.62 &0.09 &      \\
-33.2 &  17.22800 & 0.00003   & 07.9450  &0.0001 &   38.77 &0.06 &      \\
-34.5 &  17.22906 & 0.00032   & 07.9280  &0.0008 &   5.94 & 0.05 &      \\
-35.8 &  17.21846 & 0.01244   & 07.5844  &0.0204 &   8.60 & 0.31 &      \\
-37.1 &  17.22580 & 0.00115   & 07.6553  &0.0046 &   1.58 & 0.06 &      \\
-50.3 &  17.22169 & 0.00005   & 07.8103  &0.0002 &   2.93 & 0.01 &      \\
-51.6 &  17.22173 & 0.00002   & 07.8109  &0.0001 &   10.09 &0.02 &      \\
-52.9 &  17.22173 & 0.00002   & 07.8115  &0.0001 &   9.51 & 0.02 &      \\
-54.2 &  17.22169 & 0.00006   & 07.8124  &0.0003 &   2.27 & 0.01 &      \\\hline
-29.2 &  17.20965 & 0.00002   & 07.3098  &0.0000 &   85.18 & 0.09 & Jet  \\
-31.8 &  17.21571 & 0.00013   & 07.8589  &0.0005 &   15.11 & 0.10 &     \\
-33.2 &  17.21588 & 0.00004   & 07.8572  &0.0002 &   25.63 & 0.06 &     \\
-34.5 &  17.21453 & 0.00004   & 07.7631  &0.0002 &   31.66 & 0.05 &     \\
-35.8 &  17.21287 & 0.00002   & 07.6567  &0.0001 &   39.31 & 0.05 &     \\
-37.1 &  17.21284 & 0.00002   & 07.6550  &0.0001 &   30.19 & 0.04 &     \\
-38.4 &  17.21275 & 0.00006   & 07.6365  &0.0003 &   6.50 & 0.02  &    \\\hline
-50.3 &  17.22121 & 0.00005   & 07.8130  &0.0002 & 3.39 & 0.01  & Jet   \\
-51.6 &  17.22114 & 0.00003   & 07.8134  &0.0001 & 11.59 & 0.02 &     \\
-52.9 &  17.22113 & 0.00003   & 07.8135  &0.0001 & 10.93 & 0.02 &     \\
-54.2 &  17.22112 & 0.00006   & 07.8135  &0.0003 & 2.58 & 0.01  &    \\\hline
-25.3 &  17.21011 & 0.00009   & 07.3292  &0.0004 & 2.14 & 0.01  & Jet (\textbf{g1})   \\
-26.6 &  17.20985 & 0.00002   & 07.3261  &0.0001 & 17.31 & 0.03 &     \\
-27.9 &  17.20958 & 0.00002   & 07.3286  &0.0001 & 52.47 & 0.07 &     \\
-29.2 &  17.20936 & 0.00002   & 07.3309  &0.0001 & 69.73 & 0.09 &     \\
-30.5 &  17.20909 & 0.00004   & 07.3316  &0.0002 & 38.43 & 0.08 &    \\
-31.8$^b$ &  17.20338 & 0.00095   & 07.3008  &0.0017 & 13.47 & 0.19 &     \\
-33.2$^b$ &  17.20076 & 0.00020   & 07.3157  &0.0008 & 6.89 & 0.07  &    \\
-34.5$^b$ &  17.21109 & 0.00066   & 07.4804  &0.0013 & 38.92 & 0.18 &     \\
-38.4$^b$ &  17.21141 & 0.00005   & 07.3695  &0.0003 & 7.09 & 0.02  &   \\
-39.7 &  17.21092 & 0.00006   & 07.3200  &0.0003 & 3.73 & 0.01  &  \\
-41.1 &  17.21093 & 0.00011   & 07.3250  &0.0005 & 2.38 & 0.02  &    \\
-42.4 &  17.21091 & 0.00009   & 07.3224  &0.0004 & 2.27 & 0.01  &    \\
-43.7 &  17.21095 & 0.00007   & 07.3254  &0.0003 & 2.39 & 0.01  &   \\
-45.0 &  17.21099 & 0.00010   & 07.3241  &0.0004 & 1.48 & 0.01  &    \\
-46.3 &  17.21117 & 0.00019   & 07.3236  &0.0008 & 0.813 & 0.01 &     \\
-47.6 &  17.21156 & 0.00016   & 07.3175  &0.0006 & 1.08 & 0.01  &    \\
-48.9 &  17.21165 & 0.00011   & 07.3164  &0.0005 & 1.56 & 0.01  &    \\
-50.3 &  17.21143 & 0.00015   & 07.3147  &0.0006 & 1.49 & 0.01  &    \\
-51.6 &  17.21162 & 0.00058   & 07.3019  &0.0021 & 1.07 & 0.03  &    \\
-52.9 &  17.21375 & 0.00280   & 07.2891  &0.0081 & 0.989 & 0.05 &     \\
-54.2 &  17.21159 & 0.00104   & 07.3020  &0.0041 & 0.294 & 0.01 &    \\\hline
-30.5$^b$ &  17.20013 & 0.00090   & 07.2632  &0.0030 & 8.05 & 0.14  & Jet (\textbf{g2})\\
-31.8 &  17.19417 & 0.00070   & 07.0484  &0.0025 & 6.58 & 0.14  &    \\
-33.2 &  17.19286 & 0.00041   & 07.0098  &0.0016 & 4.86 & 0.08  &  \\
-34.5 &  17.19396 & 0.00074   & 07.0338  &0.0027 & 2.58 & 0.06  &   \\
-35.8 &  17.19256 & 0.00073   & 06.9989  &0.0030 & 1.69 & 0.06  &    \\
-37.1 &  17.19132 & 0.00029   & 06.9885  &0.0012 & 2.60 & 0.04  &    \\
-38.4 &  17.19088 & 0.00007   & 06.9862  &0.0003 & 3.60 & 0.02  &    \\
-39.7 &  17.19097 & 0.00005   & 06.9853  &0.0002 & 4.63 & 0.01  &   \\
-41.1 &  17.19144 & 0.00006   & 06.9828  &0.0002 & 4.49 & 0.02  &   \\
-42.4 &  17.19186 & 0.00008   & 06.9827  &0.0004 & 2.42 & 0.01  &    \\ \hline
\enddata
\tablenotetext{a}{Radio continuum source (from Rodr\'\i guez
\etal~ 2008) associated with the group of masers.} 
\tablenotetext{b}{Spot not forming part of the compact structures discussed
in the text.}
\end{deluxetable}

\clearpage

\begin{table}
\begin{center}
\scriptsize
\small
\caption{Continuum components at 1.3 mm\label{1mmcont}}
\begin{tabular}{lcccc}
\tableline\tableline
 &\multicolumn{2}{c}{Position$^a$} & Total Flux
&  \\
\cline{2-3}
Component &  $\alpha$(J2000) & $\delta$(J2000) & Density (mJy) &
Deconvolved Angular Size$^b$  \\
\tableline
West & 17.080$\pm$0.006 & 07.47$\pm$0.12 & 220$\pm$35 &
$\leq 2\rlap.{''}2 \times \leq 1\rlap.{''}5 ;~+45^\circ
\pm 29^\circ$  \\
East & 17.247$\pm$0.002 & 08.09$\pm$0.04 & 810$\pm$37 &
$1\rlap.{''}34\pm 0\rlap.{''}10 \times 0\rlap.{''}84\pm 0\rlap.{''}33;~+107^\circ
\pm 17^\circ$  \\%
\tableline
\end{tabular}
\tablenotetext{a}{Units of right
ascension are seconds with respect to 16$^h$ 58$^m$
and units of declination are arcseconds with respect to $-$42$^\circ$ 52'.}
\tablenotetext{b}{The values given are major axis $\times$ minor axis; 
position angle of major axis.
For the east component we only obtained
upper limits to the angular size.}
\end{center}
\end{table}

\clearpage

\begin{deluxetable}{rrrrr}
\tablecolumns{5}
\tablewidth{0pc}
\tablecaption{Positions of the emission peak in each velocity channel of the
SO$_2$ data.\label{so2table}}

\tablehead{

\colhead{V$_{\mbox{\scriptsize LSR}}$} &    \colhead{$\alpha$(J2000)}   & 
\colhead{$\Delta \alpha$}&
\colhead{$\delta$(J2000)} & \colhead{$\Delta \delta$}  \\ 
\colhead{(km s$^{-1}$)} &   \colhead{$16^h~ 58^m$}    & \colhead{(sec)} &  
\colhead{$-42^{\circ}~ 52'$}   & \colhead{(arcsec)}  
}
\startdata
-34.8	  &  17.294    &    0.013   & 07.88    &    0.16  \\   
-33.7	  &  17.270    &    0.004   & 08.04    &    0.08  \\  
-32.6	  &  17.268    &    0.005   & 08.11    &    0.09  \\  
-31.6	  &  17.253    &    0.003   & 07.82    &    0.05  \\ 
-30.5	  &  17.248    &    0.002   & 07.95    &    0.04  \\ 
-29.4	  &  17.222    &    0.003   & 08.02    &    0.05  \\  
-28.3	  &  17.221    &    0.003   & 07.95    &    0.06  \\ 
-27.3	  &  17.217    &    0.004   & 07.74    &    0.08  \\  
-26.2	  &  17.189    &    0.007   & 07.86    &    0.14  \\\hline   
\enddata
\end{deluxetable}

\clearpage

\end{document}